\documentclass[proceedings]{stacs}
\stacsheading{2009}{673--684}{Freiburg}
\firstpageno{673}

\theoremstyle{definition}
\theoremstyle{plain}

\usepackage{times}
\usepackage{graphicx}
\usepackage{latexsym}
\usepackage{amsmath,amssymb,stmaryrd}

\newcommand{\mi}[1]{\mathit{#1}}

\newlength{\croutw}
\newlength{\crouth}
\newcommand{\crossout}[1]%
        {\settowidth{\croutw}{$#1$}\settoheight{\crouth}{$#1$}#1%
        \hspace{-1.0\croutw}\raisebox{0.3\crouth}{\rule{\croutw}{0.1ex}}}

\newcommand{\commentout}[1]{\ignorespaces}

\newcommand{\Pow}{\mathcal{P}}

\newcommand{\bit}{\begin{itemize}}
\newcommand{\eit}{\end{itemize}}

        {\begin{enumerate}}%
        {\end{enumerate}}
        {\begin{enumerate}}%
        {\end{enumerate}}
        {\begin{enumerate}}%
        {\end{enumerate}}
        {\begin{enumerate}}%
        {\end{enumerate}}

\newcommand{\Cat}{\mathbf}

\newcommand{\Op}{{op}}

\newcommand{\impl}{\Rightarrow}

\newcommand{\Set}{\Cat{Set}}

\newcommand{\Sem}[1]{{[\![#1]\!]}}

\renewcommand{\conj}{\wedge}
\newcommand{\disj}{\vee}
\newcommand{\modimpl}{\to}
\newcommand{\modiff}{\leftrightarrow}

\newcommand{\entails}{\vdash}

        {\smallskip\par\noindent\qquad$\begin{array}{@{}l@{{}={}}l}}%
        {\end{array}$\par\noindent}

\newlength{\myboxwidth}
	{\end{center}\end{figure}}

\newcommand{\Lang}{\mathcal{L}}	
\newcommand{\FLang}{\mathcal{F}}

\newcommand{\Nat}{{\mathbb{N}}}
\newcommand{\Int}{{\mathbb{Z}}}

\newcommand{\Bag}{\mathcal{B}}
\newcommand{\Baginfty}{\mathcal{B}_\infty}
\newcommand{\Prop}{\mathsf{Prop}}

\newcommand{\contrapow}{\mathcal{Q}}

\DeclareMathOperator{\mge}{\ge}

\newcommand{\CK}{\mathit{CK}}

\newcommand{\Ax}{\mathcal{A}}

\newcommand{\CB}{\mathcal{B}}

\newenvironment{axarraycomment}%
	       {\begin{array}{@{\hspace{2em}}p{5em}p{20em}p{20em}}}{\end{array}}

\newcommand{\invlim}{\varprojlim}
\newcommand{\modelsOS}{\models^1}

\newcommand{\FA}{\mathfrak{A}}
\newcommand{\RBox}{\mathcal{R}}

\newcommand{\ALCQ}{\mathcal{ALCQ}}

\newcommand{\Lor}{\bigvee}
\newcommand{\CondArrow}{\Rightarrow}

\newcommand{\Sel}{\mathcal{S}}
\newcommand{\supp}{\mathrm{supp}}
\newcommand{\Dist}{\mathcal{D}}
\newcommand{\AtProp}{P}
\newcommand{\Form}{\FLang}

\newcommand{\inv}{^{-1}}
\newcommand{\To}{\Rightarrow}
\newcommand{\lsem}{\llbracket}
\newcommand{\rsem}{\rrbracket}
\newcommand{\biimpl}{\leftrightarrow}
\newcommand{\GML}{\mathrm{GML}}

\newcommand{\CL}{\mathrm{CL}}
\newcommand{\PML}{\mathrm{PML}}

\newcounter{blubber}

\newenvironment{sparenumerate}
{\begin{list}
  {\arabic{blubber}.}
  {\usecounter{blubber}
   \setlength{\leftmargin}{0pt}
    \setlength{\parsep}{0pt}
    \setlength{\itemindent}{3ex}
    \setlength{\itemsep}{2pt}   
    \setlength{\listparindent}{3ex}
  }
}
{\end{list}}

\newenvironment{sparitemize}
{\begin{list}{$\bullet$}{
    \setlength{\leftmargin}{0pt}
    \setlength{\parsep}{0pt}
    \setlength{\itemindent}{4ex}
    \setlength{\itemsep}{0pt}
  }
}{\end{list}}

\begin{document}

\title{Strong Completeness of Coalgebraic Modal Logics}

\author[DFKIUHB]{L. Schr{\"o}der}{Lutz
  Schr{\"o}der}
\address[DFKIUHB]{DFKI Bremen and Department of Computer Science,  
Universit\"at  Bremen}
\email{Lutz.Schroeder@dfki.de}
\author[IC]{D. Pattinson}{Dirk Pattinson}
\address[IC]{Department of Computing, 
Imperial College London}

\email{dirk@doc.ic.ac.uk}

\thanks{Work of the first author performed as part of the DFG project
  \emph{Generic Algorithms and Complexity Bounds in Coalgebraic Modal
    Logic} (SCHR 1118/5-1). Work of the second author partially
  supported by EPSRC grant EP/F031173/1}



\begin{abstract}
  Canonical models are of central importance in modal logic, in
  particular as they witness strong completeness and hence
  compactness. While the canonical model construction is well
  understood for Kripke semantics, non-normal modal logics often
  present subtle difficulties -- up to the point that canonical models
  may fail to exist, as is the case e.g.\ in most probabilistic
  logics.  Here, we present a generic canonical model construction in
  the semantic framework of coalgebraic modal logic, which pinpoints
	coherence conditions between syntax and semantics of modal logics
	that guarantee strong completeness.
  We apply this method to reconstruct canonical
  model theorems that are either known or folklore, and moreover instantiate 
  our method to obtain new strong
  completeness results. In particular, we prove strong completeness of
  graded modal logic with finite multiplicities, and of the modal
  logic of exact probabilities.
\end{abstract}

\maketitle


\noindent In modal logic, completeness proofs come in two flavours:
\emph{weak} completeness, i.e.\ derivability of all universally
valid formulas,
is often proved using \emph{finite model} constructions, and
\emph{strong} completeness, which additionally allows for a possibly
infinite set of assumptions. The latter entails recursive
enumerability of the set of consequences of a recursively enumerable
set of assumptions, and is usually established using (infinite)
\emph{canonical models}. The appeal of the first method is that it
typically entails decidability.  The second method yields a stronger
result and has some advantages of its own. First, it applies in some
cases where finite models fail to exist, which often means that the
logic at hand is undecidable. In such cases, a completeness proof via
canonical models will at least salvage recursive
enumerability. Second, it allows for schematic axiomatisations, e.g.\
pertaining to the infinite evolution of a system or to observational
equivalence, i.e.\ statements to the effect that certain states cannot
be distinguished by any formula.

In the realm of Kripke semantics, canonical models exist for a large
variety of logics and are well understood, see e.g.\
\cite{BlackburnEA01}. But there is more to modal logic than Kripke
semantics, and indeed the natural semantic structures used to
interpret a  large
class of modal logics go beyond pure relations. This includes e.g.\
the selection function semantics of conditional logics
\cite{Chellas80}, the semantics of probabilistic logics in terms of
probability distributions, and the game frame semantics of coalition 
logic~\cite{Pauly02}.  
To date, there is very little research that provides systematic
criteria, or at least a methodology, for establishing strong
completeness for logics not amenable to Kripke semantics.  This is made worse
as the question of strong completeness crucially depends on the chosen
semantic domain, which as illustrated above may differ widely. It is
precisely this variety in semantics that makes it hard to employ the
strong-completeness-via-canonicity approach, as in many cases there is
no readily available notion of canonical model.  The present work
improves on this situation by providing a widely applicable generic
canonical model construction. More precisely,  we establish the
existence of quasi-canonical models, that is, models based on the
set of maximally consistent sets of formulas that satisfy the truth
lemma, as there may be no unique, or canonical, such model in our
more general case.
In order to cover the large span of semantic structures, we avoid a
commitment to a particular class of models, and instead work within
the framework of coalgebraic modal logic~\cite{Pattinson03} which
precisely provides us with a semantic umbrella for all of the examples
above.  This is achieved by using coalgebras for an endofunctor $T$ as
the semantic domain for modal languages. As we illustrate in examples,
the semantics of particular logics is then obtained by particular
choices of~$T$. Coalgebraic modal logic serves in particular as a
general semantic framework for non-normal modal
logics. As such, it improves on neighbourhood semantics in that it
retains the full semantic structure of the original models
(neighbourhood semantics offers only very little actual semantic
structure, and in fact may be regarded as constructed from syntactic
material~\cite{SchroderPattinson07mcs}).


In this setting, our criterion can be formulated as a set of coherence
conditions that relate the syntactic component of a logic to its
coalgebraic semantics, together with a purely semantic condition
stating that the endofunctor~$T$ that defines the semantics needs to
preserve inverse limits weakly, and thus allows for a passage from the
finite to the infinite. We are initially concerned with the existence
of quasi-canonical models relative to the class of \emph{all}
$T$-coalgebras, that is, whith logics that are axiomatisable by
formulas of modal depth uniformly equal to one \cite{Schroder07}.  As
in the classical theory, the corresponding result for logics with
extra frame conditions requires that the logic is canonical, i.e. the
frame that underlies a quasi-canonical model satisfies the frame
conditions, which holds in most cases, but for the time being needs to
be established individually for each logic.

Our new criterion is then used to obtain both previously known and
novel strong completeness results. In addition to positive results, we
dissect a number of logics for which strong completeness fails and
show which assumption of our criterion is violated. In particular,
this provides a handle on adjusting either the syntax or the semantics
of the logic at hand to achieve strong completeness.
For example, we demonstrate that the failure of strong
completeness for probabilistic modal logic (witnessed e.g.\ by the set
of formulas assigning probability $\mge 1-1/n$ to an event for all $n$
but excluding probability $1$) disappears in the logic of exact
probabilities.  Moreover, we show that graded modal logic, and more
generally any description logic~\cite{BaaderEA03} with qualified
number restrictions, role hierarchies, and reflexive, transitive, and
symmetric roles, is strongly complete over the multigraph model of
\cite{DAgostinoVisser02}, which admits infinite multiplicities. While
strong completeness fails for the naive restriction of this model to
multigraphs allowing only finite multiplicities, we show how to
salvage strong completeness using additive
\mbox{(finite-)integer}-valued measures. Finally, we prove strong
completeness of several conditional logics w.r.t.\ conditional frames
(also known as selection function models); for at least one of these
logics, strong completeness was previously unknown.


\section{Preliminaries and Notation}

\noindent Our treatment of strong completeness is parametric in both
the syntax and the semantics of a wide range of modal logics. On the
syntactic side, we fix a \emph{modal similarity type} $\Lambda$
consisting of modal operators with associated arities. Given a
similarity type $\Lambda$ and a countable set $\AtProp$ of
atomic propositions, the set $\FLang(\Lambda)$ of
\emph{$\Lambda$-formulas} is inductively defined by the grammar
\begin{equation*}
 \Form(\Lambda) \ni \phi, \psi ::= p \mid \bot \mid \neg\phi\mid\phi \conj \psi \mid
L(\phi_1, \dots, \phi_n) 
\end{equation*}
where $p \in \AtProp$ and $L \in \Lambda$ is $n$-ary; further boolean
operators ($\disj$, $\modimpl$, $\modiff$, $\top$) are defined as
usual.  Given any set $X$ (e.g.\ of formulas, atomic propositions, or
sets (!)), we write $\Prop(X)$ for the set of propositional formulas
over $X$ and $\Lambda(X) = \lbrace L(x_1, \dots, x_n) \mid L \in
\Lambda \mbox{ is $n$-ary}, x_1, \dots, x_n \in X \rbrace$ for the set
of formulas arising by applying exactly one operator to elements of
$X$.  We instantiate our results to a variety of settings later with
the following similarity types:
\begin{exas} \label{expl:sim-types}
\begin{sparenumerate}
\item The similarity type $\Lambda_K$ of standard modal logic consists of a
single unary operator $\Box$.
\item Conditional logic \cite{Chellas80} is defined over the
  similarity type $\Lambda_\CL = \lbrace \To \rbrace$ where the binary
  operator $\To$ is read as a non-monotonic conditional (default,
  relevant etc.), usually written in infix notation.
\item Graded modal operators~\cite{Fine72} appear in expressive
  description logics~\cite{BaaderEA03} in the guise of so-called
  qualified number restrictions; although we discuss only modal
  aspects, we use mostly description logic notation and terminology
  below. The operators of graded modal logic (GML) are $\Lambda_{\GML}
  = \lbrace (\mge k) \mid k \in \Nat \rbrace$ with $(\mge k)$ unary.
  We write $\mge k.\, \phi$ instead of $(\mge k) \phi$. A formula
  $\mge k.\, \phi$ is read as `at least $k$ successor states satisfy
  $\phi$', and we abreviate $\Box \phi = \neg \mge 1. \neg \phi$.
\item The similarity type $\Lambda_\PML$ of probabilistic modal logic
  (PML)~\cite{LarsenSkou91} contains the unary modal operators $L_p$
  for $p \in \mathbb{Q} \cap [0, 1]$, read as `with probability at
  least $p$, \dots'.
\end{sparenumerate}
\end{exas}
\noindent We split axiomatisations of modal logics into two parts: the
first group of axioms is responsible for axiomatising the logic
w.r.t. the class of \emph{all} (coalgebraic) models, whereas the
second consists of frame conditions that impose additional conditions
on models. As the class of all coalgebraic models, introduced below,
can always be axiomatised by formulas of \emph{rank $1$}, i.e.\
containing exactly one level of modal operators~\cite{Schroder07} (and
conversely, every collection of such axioms admits a complete
coalgebraic semantics~\cite{SchroderPattinson07mcs}), we restrict the
axioms in the first group accordingly.  More formally:
\begin{defi}
  A \emph{(modal) logic} is a triple $\Lang=(\Lambda, \Ax, \Theta)$
  where $\Lambda$ is a similarity type, $\Ax \subseteq
  \Prop(\Lambda(\Prop(\AtProp)))$ is a set of \emph{rank-1 axioms},
  and $\Theta \subseteq \Form(\Lambda)$ is a set of \emph{frame
    conditions}. We say that $\Lang$ is a \emph{rank-1 logic} if
  $\Theta=\emptyset$. If $\phi \in \Form(\Lambda)$, we write
  $\entails_{\Lang} \phi$ if $\phi$ can be derived from $\Ax \cup
  \Theta$ with the help of propositional reasoning, uniform
  substitution, and the congruence rule: from $\phi_1 \biimpl \psi_1,
  \dots, \phi_n \biimpl \psi_n$ infer $L(\phi_1, \dots, \phi_n)
  \biimpl L(\psi_1, \dots, \psi_n)$ whenever $L \in \Lambda$ is
  $n$-ary. For a set $\Phi \subseteq \Form(\Lambda)$ of assumptions,
  we write $\Phi \entails_{\Lang} \phi$ if $\entails_{\Lang} \phi_1
  \land \dots \land \phi_n \to \phi$ for (finitely many) $\phi_1,
  \dots, \phi_n \in \Phi$. A set $\Phi$ is \emph{$\Lang$-inconsistent}
  if $\Phi\entails_\Lang\bot$, and otherwise
  \emph{$\Lang$-consistent}.
\end{defi}
%

%
\begin{exas}\label{expl:axioms}
\begin{sparenumerate}
\item \label{item:ax-kripke} The modal logic $K$ comes about as the
  rank-1 logic $(\Lambda_K, \Ax_K, \emptyset)$ where $\Ax_k = \lbrace
  \Box\top,\Box(p \to q) \to (\Box p \to \Box q) \rbrace$. The logics
  $K4, S4, KB, \dots$ arise as $(\Lambda_K, \Ax_K, \Theta)$ where
  $\Theta$ contains the additional axioms that define the respective
  logic~\cite{BlackburnEA01}, e.g.\ $\Theta=\{\Box p\to\Box\Box p\}$
  in the case of $K4$.
\item\label{item:ax-cond} For conditional logic, we take the
  similarity type $\Lambda_\CL$ together with rank-1 axioms $r\To
  \top$, $r \To (p \to q) \to ((r \To p) \to (r \To q))$ stating that
  the binary conditional is normal in its second argument. Typical
  additional rank-1 axioms are
  \begin{equation*}
    \begin{axarraycomment}
      \textrm{(ID)} & $a\To a$ &\emph{(identity)}\\
      \textrm{(DIS)} & $(a\To c) \conj (b\To c)\to ((a\disj b)\To c)$&
      \emph{(disjunction)}\\
      \textrm{(CM)} & $(a\To c) \conj (a\To b)\to ((a\conj b)\To c)$&
      \emph{(cautious monotony)}\\
    \end{axarraycomment}
  \end{equation*}
  which together form the so-called \emph{System C}, a modal version of the
  well-known KLM (Krauss/Lehmannn/Magidor) axioms of default reasoning
  due to Burgess~\cite{Burgess81}.
\item\label{item:ax-gml} The axiomatisation of GML
  given in~\cite{Fine72} consists of the rank-1 axioms
\begin{itemize}
\item[] $\Box (p \to q) \to (\Box p \to \Box q)$
\item[] $\mge k.\, p \to \mge l.\,p$ for $l < k$
\item[] $\mge k.\, p \biimpl \Lor_{i=0, \dots, k} \mge i.\, (p \land q)
\land \mge(k-i).\, (p \land \neg q)$
\item[] $\Box (p \to q) \to (\mge k.\,p \to \mge k.\,q)$
\end{itemize}
Frame conditions of interest include e.g. reflexivity ($p\modimpl\mge
1.\,p$), symmetry ($p\modimpl\Box \,\mge 1.  \,p$), and transitivity
($\mge 1.\,\mge n.\,p\modimpl\mge n.\,p$).
\end{sparenumerate}
\end{exas}
%
%
%
%
\noindent To keep our results parametric also in the semantics of
modal logic, we work in the framework of \emph{coalgebraic modal
  logic} in order to achieve a uniform and coherent presentation.  In
this framework, the particular shape of models is encapsulated by an
endofunctor $T: \Set \to \Set$, the \emph{signature functor} (recall
that such a functor maps every set $X$ to a set $TX$, and every map
$f:X\to Y$ to a map $Tf:TX\to TY$ in such a way that composition and
identities are preserved), which may be thought of as a parametrised
data type. We fix the data $\Lambda$, $\Lang$, $T$ etc.\ throughout
the generic part of the development. The role of models in then played
by $T$-coalgebras:
\begin{defi}
A \emph{$T$-coalgebra} is a pair $\mathbb{C} = (C, \gamma)$ where $C$ is a set
(the \emph{state space} of $\mathbb{C}$)
and $\gamma: C \to  T C$ is a
function, the transition structure of $\mathbb{C}$.
\end{defi}
\noindent We think of $TC$ as a type of successors,
polymorphic in $C$. The
transition structure $\gamma$
associates a structured collection of successors 
$\gamma(c)$ to each state $x\in C$.
%
%
\noindent The following choices of signature functors give rise to the
semantics of the modal logics discussed in Expl.
\ref{expl:axioms}.
\begin{exas}\label{expl:coalgml}
\begin{sparenumerate}
\item \label{item:Kripke} Coalgebras for the covariant powerset
  functor $\Pow$ defined on sets $X$ by $\Pow(X) = \lbrace A \mid A
  \subseteq X \rbrace$ and on maps $f$ by $\Pow(f)(A)=f[A]$ are Kripke
  frames, as relations $R \subseteq W \times W$ on a set $W$ of worlds
  are in bijection with functions of type $W \to \Pow(W)$.
  Restricting the powerset functor to \emph{finite} subsets,
  i.e. putting $\Pow_\omega(X) = \lbrace A \subseteq X \mid A \mbox{
    finite} \rbrace$, one obtains the class of image finite Kripke
  frames as $\Pow_{\omega}$-coalgebras. 
\item\label{item:cond} The semantics of conditional logic is captured
  coalgebraically by the endofunctor $\Sel$ that maps a set $X$ to the
  set $(\Pow(X) \to \Pow(X))$ of selection functions over $X$ (the
  action of $\Sel$ on functions $f: X \to Y$ is given by
  $\Sel(f)(s)(B) = f [ s(f\inv[B])]$).  The ensuing $\Sel$-coalgebras
  are precisely the conditional frames of \cite{Chellas80}.
\item\label{item:gml} The \emph{(infinite) multiset functor}
  $\Baginfty$ maps a set $X$ to the set $\Baginfty X$ of multisets
  over $X$, i.e.\ functions of type $X \to \Nat \cup \lbrace \infty
  \rbrace$.
  Accordingly, $\Baginfty$-coalgebras are \emph{multigraphs} (graphs
  with edges annotated by multiplicities). Multigraphs provide an
  alternative semantics for GML which is in many
  respects more natural than the original Kripke
  semantics~\cite{DAgostinoVisser02}, as also confirmed by new
  results below.
\item\label{item:pml} Finally, if $\supp(\mu) = \lbrace x \in X \mid
  \mu(x) \neq 0 \rbrace$ is the support of a function $\mu: X \to [0,
  1]$ and $\Dist(X) = \lbrace \mu: X \to [0, 1] \mid \supp(\mu) \mbox{
    finite}, \sum_{x \in X} \mu(x) = 1 \rbrace$ is the set of finitely
  supported probability distributions on $X$, then $\Dist$-coagebras
  are probabilistic transition systems, the semantic domain of
  PML.
\end{sparenumerate}
\end{exas}
\noindent The link between coalgebras and modal languages is provided
by predicate liftings~\cite{Pattinson03}, which are used to interpret
modal operators. Essentially, predicate liftings convert predicates on
the state space $X$ into predicates on the set $TX$ of structured
collections of states:
\begin{defi}\label{def:lifting}\cite{Pattinson03}
  An \emph{$n$-ary predicate lifting} ($n\in\Nat$) for $T$
  is a family of maps $\lambda_X:\Pow{X}^n \to\Pow{TX}$, where $X$
  ranges over all sets, satisfying the \emph{naturality} condition
  \begin{equation*}
    \lambda_X(f^{-1}[A_1],\dots,f^{-1}[A_n])=(Tf)^{-1}[\lambda_Y(A_1,\dots,A_n)]
  \end{equation*}
  for all $f:X\to Y$, $A_1,\dots,A_n\in\Pow{Y}$. (For the
  categorically minded, $\lambda$ is a natural transformation
  $\contrapow^n\to\contrapow\circ T^\Op$, where $\contrapow$ denotes
  contravariant powerset.) A \emph{structure} for a similarity type
  $\Lambda$ over an endofunctor $T$ is the assignment of an $n$-ary
  predicate lifting $\lsem L \rsem$ to every $n$-ary modal operator $L
  \in \Lambda$.
\end{defi}\noindent
\noindent Given a valuation $V: \AtProp \to \Pow(C)$ of the
propositional variables and a $T$-coalgebra $(C, \gamma)$, a structure
for $\Lambda$ allows us to define a satisfaction relation
$\models_{(C, \gamma,V)}$ between states of $C$ and formulas $\phi \in
\Form(\Lambda)$ by stipulating that $c\models_{(C, \gamma,V)}p$ iff
$c\in V(p)$ and
\[
c\models_{(C, \gamma, V)} L (\phi_1, \dots, \phi_n)\;\textrm{ iff }\;
\gamma(c)\in \lsem L \rsem_C (\lsem \phi_1 \rsem,
\dots, \lsem \phi_n \rsem),
\]
where $\lsem \phi \rsem=\{c\in C\mid c\models_{(C, \gamma,V)}\phi\}$.
An \emph{$\Lang$-model} is now a \emph{model}, i.e.\ a triple $(C,
\gamma, V)$ as above, such that $c \models_{(C, \gamma,V)} \psi$ for
all all $c\in C$ and all substitution instances $\psi$ of $\Ax \cup
\Theta$.  An \emph{$\Lang$-frame} is a $T$-coalgebra $(C, \gamma)$ such that
$(C, \gamma, V)$ is an $\Lang$-model for all valuations~$V$.  The
reader is invited to check that the following predicate liftings
induce the standard semantics for the modal languages introduced in
Expl. \ref{expl:sim-types}.
\begin{exas}\label{expl:structure}
\begin{sparenumerate}
\item A structure for $\Lambda_K$ over the covariant powerset functor
  $\Pow$ is given by $\lsem \Box \rsem_X (A) = \lbrace Y \in \Pow(X)
  \mid Y \subseteq A \rbrace$. The frame classes defined by the frame
  conditions mentioned in Expl.~\ref{expl:axioms}.\ref{item:ax-kripke}
  are well-known; e.g.\ a Kripke frame $(X,R)$ is a $K4$-frame iff $R$
  is transitive.
\item Putting $\lsem \To \rsem_X (A, B) = \lbrace f \in \Sel(X) \mid
  f(A) \subseteq B \rbrace$ reconstructs the semantics of conditional
  logic in a coalgebraic setting. 
\item\label{item:gml-struct} A structure for GML over
  $\Baginfty$ is given by $\lsem (\mge k) \rsem_X (A) = \lbrace f: X
  \to \Nat \cup \lbrace \infty \rbrace \mid \sum_{x \in A} f(x) \geq
	k \rbrace$. The frame conditions mentioned in
  Expl.~\ref{expl:axioms}.\ref{item:ax-gml} correspond to conditions
  on multigraphs that can be read off directly from the logical
  axioms. E.g.\ a multigraph satisfies the transitivity axiom $\mge
  1.\,\mge n.\,p\modimpl\mge n.\,p$ iff whenever $x$ has non-zero
  transition multiplicity to $y$ and $y$ has transition multiplicity
  at least $n$ to $z$, then $x$ has transition multiplicity at least
  $n$ to $z$.
\item The structure over $\Dist$ that captures PML coalgebraically is
  given by the the predicate lifting $\lsem L_p \rsem_X (A) = \lbrace
  \mu \in \Dist(X) \mid \sum_{x \in A} \mu(x) \geq p \rbrace$ for $p
  \in [0, 1 ] \cap \mathbb{Q}$.
\end{sparenumerate}
\end{exas}
\noindent From now on, \emph{fix a modal logic $\Lang=(\Lambda, \Ax,
  \Theta)$ and a structure for $\Lambda$ over a functor~$T$}. 
We say that $\Lang$ is \emph{strongly complete} for some class of
models if every $\Lang$-consistent set of formulas is satisfiable in
some state of some model in that class. Restricting to \emph{finite}
sets $\Phi$ defines the notion of \emph{weak completeness}; many
coalgebraic modal logics are only weakly complete~\cite{Schroder07}.
\begin{defi}
  Let $X$ be a set. If $\psi \in \FLang(\Lambda)$ and $\tau: \AtProp
  \to \Pow(X)$ is a valuation, we write $\psi \tau$ for the result of
  substituting $\tau(p)$ for $p$ in $\psi$, with propositional
  subformulas evaluated according to the boolean algebra structure of
  $\Pow(X)$. (Hence, $\psi\tau$ is a formula over the set $\Pow(X)$ of
  atoms.)  A formula $\phi \in \Prop(\Lambda(\Pow(X))$ is
  \emph{one-step $\Lang$-derivable}, denoted $\entails^1_\Lang \phi$,
  if $\phi$ is propositonally entailed by the set $ \lbrace \psi \tau
  \mid \tau:\AtProp \to \Pow(X), \psi \in \Ax \rbrace$. A set $\Phi
  \subseteq \Prop(\Lambda(\Pow(X)))$ is \emph{one-step
    $\Lang$-consistent} if there do not exist formulas $\phi_1, \dots,
  \phi_n \in \Phi$ such that $\entails_\Lang^1
  \neg(\phi_1\land\dots\land\phi_n)$. Dually, the \emph{one-step
    semantics} $\lsem \phi \rsem_X^1 \subseteq T X$ of a formula $\phi
  \in \Prop(\Lambda(\Pow(X))$ is defined inductively by $\lsem L(A_1,
  \dots, A_n) \rsem_X^1 = \lsem L \rsem_X ( A_1, \dots, A_n)$ for
  $A_1, \dots, A_n \subseteq X$. A set $\Phi \subseteq
  \Prop(\Lambda(\Pow(X)))$ is \emph{one-step satisfiable} if
  $\bigcap_{\phi \in \Phi} \lsem \phi \rsem_X^1 \neq \emptyset$.  We
  say that $\Lang$ (or $\Lambda$) is \emph{separating} if $t\in TX$ is
  uniquely determined by the set $\{\phi\in\Lambda(\Pow(X))\mid
  t\in\Sem{\phi}^1_X\}$. We call $\Lang$ (or $\Ax$) \emph{one-step
    sound} if every one-step derivable formula $\phi \in
  \Prop(\Lambda(\Pow(X)))$ is one-step valid, i.e. $\lsem \phi
  \rsem_X^1 = X$.
\end{defi}
\noindent \emph{Henceforth, we assume that $\Lang$ is one-step sound},
so that every $T$-coalgebra satisfies the rank-1 axioms; in the
absence of frame conditions ($\Theta = \emptyset$), this means in
particular that every $T$-coalgebra is an $\Lang$-frame. The above
notions of one-step satisfiability and one-step consistency are the
main concepts employed in the proof of strong completeness in the
following section. 

Given a structure for $\Lambda$ over $T$, every set $\CB$ of rank-1
axioms over $\Lambda$ defines a subfunctor $T_\CB$ of $\CB$ with
$T_\CB(X)=\bigcap\{\Sem{\phi\tau}^1_X\mid\phi\in\CB,\tau:P\to\Pow(X)\}\subseteq
TX$. This functor induces a structure for which $\CB$ is one-step
sound.
\begin{exa}\label{expl:subfunctors}
  The additional rank-1 axioms of
  Expl.~\ref{expl:axioms}.\ref{item:ax-cond} induce subfunctors
  $\Sel_\CB$ of the functor $\Sel$ of
  Expl.~\ref{expl:coalgml}.\ref{item:cond}. E.g.\ we have
  \begin{align*}
    \Sel_{\{\mi{ID}\}}X&=\{f\in\Sel(X)\mid \forall A\subseteq X.\,
    f(A)\subseteq A\}\\
    \Sel_{\{\mi{ID,DIS}\}}X&=\{f\in\Sel(X)\mid \forall A,B\subseteq X.\,
    f(A)\subseteq A \conj f(A\cup B)\subseteq
    f(A)\cup f(B) \}\\
    \Sel_{\{\mi{ID,DIS,CM}\}}X&=
    \{f\in\Sel(X)\mid \forall A,B\subseteq X.\,
    f(A)\subseteq A \conj
    (f(B)\subseteq A \impl f(A)\cap B\subseteq f(B))\}
\end{align*}
(it is an amusing exercise to verify the last claim).
\end{exa}

\section{Strong Completeness Via Quasi-Canonical Models}

\noindent 
We wish to establish strong completeness of $\Lang$ by defining a
suitable $T$-coalgebra structure $\zeta$ on the set~$S$ of maximally
$\Lang$-consistent subsets of $\FLang(\Lambda)$, equipped with the
standard valuation $V(p)=\{\Gamma\in S\mid p\in\Gamma\}$. The crucial
property required is that $\zeta$ be \emph{coherent}, i.e.
\begin{equation*}
  \zeta(\Gamma)\in \Sem{L}(\hat\phi_1,\dots,\hat\phi_n) \iff 
  L(\phi_1,\dots,\phi_n)\in \Gamma,
\end{equation*}
where $\hat\phi=\{\Delta\in S\mid \phi\in\Delta\}$, for $L\in\Lambda$
$n$-ary, $\Gamma\in S$, and $\phi_1,\dots,\phi_n\in\FLang(\Lambda)$,
as this allows proving, by a simple induction over the structure of
formulas,
\begin{lem}[Truth lemma]
  If $\zeta$ is coherent, then for all formulas $\phi$,
  $\Gamma\models_{(S,\zeta,V)}\phi$ iff $\phi\in\Gamma$.
\end{lem}
\noindent We define a \emph{quasi-canonical model} to be a model
$(S,\zeta,V)$ with $\zeta$ coherent; the term quasi-canonical serves
to emphasise that the coherence condition does not determine the
transition structure $\zeta$ uniquely. By the truth lemma,
quasi-canonical models for $\Lang$ are $\Lang$-models, i.e.\ satisfy
all substitution instances of the frame conditions. The first question
is now under which circumstances quasi-canonical models exist; we
proceed to establish a widely applicable criterion. This criterion has
two main aspects: a \emph{local} form of strong completeness involving
only finite sets, and a preservation condition on the functor enabling
passage from finite sets to certain infinite sets. We begin with the
latter part:
\begin{defi}
  A \emph{surjective $\omega$-cochain (of finite sets)} is a sequence
  $(X_n)_{n\in\Nat}$ of (finite) sets equipped with surjective
  functions $p_n:X_{n+1}\to X_n$ called \emph{projections}. The
  \emph{inverse limit} $\invlim X_n$ of $(X_n)$ is the set
  $\{(x_i)\in\prod_{i\in\Nat} X_i \mid \forall n.\,p_n(x_{n+1})=x_n\}$ of
  \emph{coherent} families $(x_i)$. The \emph{limit projections} are
  the maps $\pi_i((x_n)_{n\in\Nat})=x_i$, $i\in\Nat$; note that the
  $\pi_i$ are surjective, i.e.\ every $x\in X_i$ can be extended to a
  coherent family. Since all set functors preserve surjections,
  $(TX_n)$ is a surjective $\omega$-cochain with projections
  $Tp_n$. The functor $T$ \emph{weakly preserves inverse limits of
    surjective $\omega$-cochains of finite sets} if for every
  surjective $\omega$-cochain $(X_n)$ of finite sets, the canonical
  map $T(\invlim X_n)\to\invlim TX_n$ is surjective, i.e.\ every
  coherent family $(t_n)$ in $\prod TX_n$ is \emph{induced} by a (not
  necessarily unique) $t\in T(\invlim X_n)$ in the sense that
  $T\pi_n(t)=t_n$ for all $n$.
\end{defi}

\begin{exa}\label{exa:cochains}
  Let $A$ be a finite alphabet; then the sets $A^n$, $n\in\Nat$, form
  a surjective $\omega$-cochain of finite sets with projections
  $p_n:A^{n+1}\to A^n$, $(a_1,\dots,a_{n+1})\mapsto
  (a_1,\dots,a_n)$. The inverse limit $\invlim A^n$ is the set
  $A^\omega$ of infinite sequences over $A$. The covariant powerset
  functor $\Pow$ preserves this inverse limit weakly: given a coherent
  family of subsets $B_n\subseteq A^n$, i.e.\ $p_n[B_{n+1}]=B_n$ for
  all $n$, we define the set $B\subseteq A^\omega$ as the set of all
  infinite sequences $(a_n)_{n\ge 1}$ such that $(a_1,\dots,a_n)\in
  B_n$ for all $n$; it is easy to check that indeed $B$ induces the
  $B_n$, i.e.\ $\pi_n[B]=B_n$. However, $B$ is by no means uniquely
  determined by this property: Observe that $B$ as just defined is a
  safety property. The intersection of $B$ with any liveness property
  $C$, e.g.\ the set $C$ of all infinite sequences containing
  infinitely many occurrences of a fixed letter in $A$, will also
  satisfy $\pi_n[B\cap C]=B_n$ for all $n$.
\end{exa}
\noindent The second part of our criterion is an infinitary version of
a local completeness property called one-step completeness, which has
been used previously in \emph{weak} completeness
proofs~\cite{Pattinson03,Schroder07}.
\begin{defi}
  We say that $\Lang$ is \emph{strongly one-step complete over finite
    sets} if for finite $X$, every one-step consistent subset $\Phi$
  of $\Prop(\Lambda(\Pow(X)))$ is one-step satisfiable.
\end{defi}
\noindent 
The difference with plain one-step completeness is that $\Phi$ above
may be infinite. Consequently, strong and plain one-step completeness
coincide in case the modal similarity type $\Lambda$ is finite, since
in this case, $\Prop(\Lambda(\Pow(X)))$ is, for finite $X$, finite up
to propositional equivalence. The announced strong completeness
criterion is now the following.
\begin{thm}\label{thm:can-model}
  If $\Lang$ is strongly one-step complete over finite sets and
  separating, $\Lambda$ is countable, and $T$ weakly preserves inverse
  limits of surjective $\omega$-cochains of finite sets, then $\Lang$
  has a quasi-canonical model.
\end{thm}
\proof[Proof sketch]
  The most natural argument is via the dual adjunction between sets
  and boolean algebras that associates to a set the boolean algebra of
  its subsets, and to a boolean algebra the set of its
  ultrafilters. For economy of presentation, we outline a direct
  proof instead: we prove that
  \begin{itemize}
  \item[($*$)] every maximally one-step consistent
    $\Phi\subseteq\Prop(\Lambda(\FA))$ is one-step satisfiable,\\ where
    $\FA=\{\hat\phi\mid\phi\in\FLang(\Lambda)\}\subseteq\Pow(S)$.
  \end{itemize}
  The existence of the required coherent coalgebra structure $\zeta$
  on $S$ follows immediately, since the coherence requirement for
  $\zeta(\Gamma)$, $\Gamma\in S$, amounts to one-step satisfaction of
  a maximally one-step consistent subset of $\Prop(\Lambda(\FA))$. 

  To prove ($*$), let $\Lambda=\{L_n\mid n\in\Nat\}$, let
  $\AtProp=\{p_n\mid n\in\Nat\}$, let $\FLang_n$ denote the set of
  $\Lambda$-formulas of modal nesting depth at most $n$ that employ
  only modal operators from $\Lambda_n=\{L_0,\dots,L_n\}$ and only the
  atomic propositions $p_0,\dots,p_n$, and let $S_n$ be the set of
  maximally consistent subsets of $\FLang_n$. Then $S$ is (isomorphic
  to) the inverse limit $\invlim S_n$, where the projections
  $S_{n+1}\to S_n$ and the limit projections $S\to S_n$ are just
  intersection with $\FLang_n$. As the sets $S_n$ are finite, we
  obtain by strong one-step completeness $t_n\in TS_n$ such that
  $t_n\modelsOS_{S_n} \Phi\cap\Prop(\Lambda(\FA_n))$, where
  $\FA_n=\{\hat\phi\cap S_n\mid\phi\in\FLang_n\}$. By separation,
  $(t_n)_{n\in\Nat}$ is coherent, and hence is induced by some $t\in
  TS$ by weak preservation of inverse limits; then,
  $t\modelsOS_S\Phi$.\qed
\noindent
Together with the Lindenbaum Lemma we obtain strong completeness as
a corollary.
\begin{cor} \label{cor:str-comp}
Under the conditions of Thm.~\ref{thm:can-model},
$\Lang$
is strongly complete for $\Lang$-models.
\end{cor}
\noindent Both Thm.~\ref{thm:can-model} and
Cor.~\ref{cor:str-comp} do apply to the case that $\Lang$ has
frame conditions. When $\Lang$ is of rank~1 (i.e.\
$\Theta=\emptyset$), Cor.~\ref{cor:str-comp} implies that $\Lang$
is strongly complete for (models based on) $\Lang$-frames.  In the
presence of frame conditions, the underlying frame of an $\Lang$-model
need not be an $\Lang$-frame, so that the question arises whether
$\Lang$ is also strongly complete for $\Lang$-frames. In applications,
positive answers to this question, usually referred to as the
canonicity problem, typically rely on a judicious choice of
quasi-canonical model to ensure that the latter is an $\Lang$-frame,
often the largest quasi-canonical model under some ordering on
$TS$. Detailed examples are given in Sec.~\ref{sec:examples}.
\begin{rem}
  It is shown in~\cite{KurzRosicky} that $T$ admits a strongly
  complete modal logic if $T$ weakly preserves (arbitrary) inverse
  limits \emph{and preserves finite sets}. The essential contribution
  of the above result is to remove the latter restriction, which fails
  in important examples. Moreover, the observation that we need only
  consider \emph{surjective} $\omega$-cochains is relevant in some
  applications, see below.
\end{rem}

\begin{rem}
  A last point that needs clearing up is whether strong completeness
  of coalgebraic modal logics can be established by some more general
  method than quasi-canonical models of the quite specific shape used
  here. The answer is negative, at least in the case of rank-1 logics
  $\Lang$: it has been shown in~\cite{KurzPattinson05} that every such
  $\Lang$ admits models which consist of the maximally
  \emph{satisfiable} sets of formulas and obey the truth lemma. Under
  strong completeness, such models are quasi-canonical.

  This seems to contradict the fact that some canonical model
  constructions in the literature, notably the canonical Kripke models
  for graded modal logics~\cite{Fine72,DeCaro88}, employ state spaces
  which have multiple copies of maximally consistent sets.  The above
  argument indicates that such logics fail to be coalgebraic, and
  indeed this is the case for GML with Kripke
  semantics. As mentioned above, GML has an alternative
  coalgebraic semantics over multigraphs, and we show below that this
  semantics does admit quasi-canonical models in our sense.
\end{rem}

\section{Examples}\label{sec:examples}

\noindent We now show how the generic results of the previous section
can be applied to obtain canonical models and associated strong
completeness and compactness theorems for a large variety of
structurally different modal logics. We have included some negative
examples where canonical models necessarily fail to exist due to
non-compactness, and we analyse which conditions of
Thm.~\ref{thm:can-model} fail in each case. We emphasise that in
the positive examples, the verification of said conditions is entirely
stereotypical. Weak preservation of inverse limits of surjective
$\omega$-cochains usually holds without the finiteness assumption,
which is therefore typically omitted.

\begin{exa}[Strong completeness of Kripke semantics for $K$]
  Recall from Expl.~\ref{expl:coalgml}.\ref{item:Kripke} that Kripke
  frames are coalgebras for the powerset functor $TX = \Pow(X)$.
  Strong completeness of $K$ with respect to Kripke semantics is, of
  course, well known.  We briefly illustrate how this can be derived
  from our coalgebraic treatment.  To see that $K$ is strongly
  one-step complete over finite sets $X$, let
  $\Phi\subseteq\Prop(\Lambda_K(\Pow(X)))$ be maximally one-step
  consistent. It is easy to check that $\{x\in
  X\mid\Diamond\{x\}\in\Phi\}$ satisfies $\Phi$.  To prove that the
  powerset functor weakly preserves inverse limits, let $(X_n)$ be an
  $\omega$-cochain, and let $(A_n\in\Pow(X_n))$ be a coherent
  family. Then $(A_n)$ is itself a cochain, and the set $A=\invlim
  A_n\subseteq\invlim X_n$ induces $(A_n)$ (w.r.t.\ the subset
  ordering on $\Pow(X)$). Separation is clear. By
  Thm.~\ref{thm:can-model}, there exists a quasi-canonical Kripke
  model for all normal modal logics.  In particular, the standard
  canonical model~\cite{Chellas80} is quasi-canonical; it witnesses
  strong completeness (w.r.t.\ frames) of all canonical logics such as
  $K4$, $S4$, $S5$.
\end{exa}

\begin{exa}[Failure of strong completeness of $K$ over finitely
branching models]
As seen in Expl.~\ref{expl:coalgml}.\ref{item:Kripke}, finitely
branching Kripke frames are coalgebras for the finite powerset functor
$\Pow_\omega$. It is clear that quasi-canonical models fail to exist
in this case, as compactness fails over finitely branching frames: one
can easily construct formulas $\phi_n$ that force a state to have at
least $n$ different successors.  The obstacle to the application of
Thm.~\ref{thm:can-model} is that the finite powerset functor fails to
preserve inverse limits weakly, as the inverse limit of an
$\omega$-cochain of finite sets may fail to be finite.
\end{exa}

\begin{exa}[Conditional logic]
  Recall from Expl.~\ref{expl:coalgml}.\ref{item:cond} that the
  conditional logic $\CK$ is interpreted over the functor
  $\Sel(X)=\Pow(X)\to\Pow(X)$. To prove strong one-step completeness
  over finite sets $X$, let $\Phi\subseteq\Prop(\Lambda_\CL(\Pow(X)))$
  be maximally one-step consistent. Define $f:\Pow(X)\to\Pow(X)$ by
  $f(A)=\bigcap\{B\subseteq X\mid A\CondArrow B\in\Phi\}$; it is
  mechanical to check that $f\modelsOS\Phi$. To see that $\Sel$ weakly
  preserves inverse limits, let $(X_n)$ be a surjective
  $\omega$-cochain, let $X=\invlim X_n$, and let $(f_n\in\Sel(X_n))$
  be coherent. Define $f:\Pow(X)\to\Pow(X)$ by letting $(x_n)\in f(A)$
  for a coherent family $(x_n)\in X$ iff whenever $A=\pi_n^{-1}[B]$
  for some $n$ and some $B\subseteq X_n$, then $x_n\in f_n(B)$. Using
  surjectivity of the projections of $(X_n)$, it is straightforward to
  prove that $f$ induces $(f_n)$. Finally, separation is clear. By
  Thm.~\ref{thm:can-model}, it follows that the conditional logic
  $\CK$ has a quasi-canonical model, and hence that $\CK$ is strongly
  complete for conditional frames.  In the case of the additional
  rank-1 axioms mentioned in
  Expl.~\ref{expl:axioms}.\ref{item:ax-cond} and the corresponding
  subfunctors of $\Sel$ described in Expl.~\ref{expl:subfunctors}, the
  situation is as follows.

  \textbf{Identity:} The functor $\Sel_{\{\mi{ID}\}}$ weakly preserves
  inverse limits of surjective $\omega$-cochains. In the notation
  above, put $(x_n)\in f(A)$ iff the condition above holds and
  $(x_n)\in A$.

  \textbf{Identity and disjunction:} The functor
  $\Sel_{\{\mi{ID},\mi{DIS}\}}$ weakly preserves inverse limits of
  surjective $\omega$-cochains: put $(x_n)\in f(A)$ iff $(x_n)\in A$
  and whenever $(x_n)\in\pi_m^{-1}B\subseteq A$, then $x_m\in f_m(B)$.

  \textbf{System C:} It is open whether the the functor
  $\Sel_{\{\mi{ID},\mi{DIS},\mi{CM}\}}$ weakly preserves inverse
  limits of surjective $\omega$-cochains, and whether System C is
  strongly complete over conditional frames.

  Indeed it appears to be an open problem to find \emph{any} semantics
  for which System C is strongly complete, other than the generalised
  neighbourhood semantics as described e.g.\
  in~\cite{SchroderPattinson07mcs}, which is strongly complete for
  very general reasons but provides little in the way of actual
  semantic information. The classical preference semantics according
  to Lewis is only known to be weakly
  complete~\cite{Burgess81}. Friedman and
  Halpern~\cite{FriedmanHalpern01} do silently prove strong
  completeness of System C w.r.t.\ plausibility measures; however, on
  close inspection the latter turn out to be essentially equivalent to
  the above-mentioned generalised neighbourhood semantics. Moreover,
  Segerberg~\cite{Segerberg89} proves strong completeness for a whole
  range of conditional logics over \emph{general} conditional frames,
  where, in analogy to corresponding terminology for Kripke frames, a
  general conditional frame is equipped with a distinguished set of
  \emph{admissible propositions} limiting both the range of valuations
  and the domain of selection functions. In contrast, our method
  yields full conditional frames in which the frame conditions hold
  for \emph{any} valuation of the propositional variables. While in
  the case of $\CK$ and its extension by $\mi{ID}$ alone, these models
  differ from Segerberg's only in that they insert default values for
  the selection function on non-admissible propositions, the
  canonical model for the extension of $\CK$ by $\{\mi{ID},\mi{DIS}\}$
  has non-trivial structure on non-admissible propositions, and we
  believe that our strong completeness result for this logic is
  genuinely new.
\end{exa}

\begin{exa}[Strong completeness of GML over
  multigraphs]\label{expl:gml-infty}
  Recall from Expl.~\ref{expl:coalgml}.\ref{item:gml} that graded
  modal logic (GML) has a coalgebraic semantics in terms of the multiset
  functor $\Baginfty$. To prove strong one-step completeness over
  finite sets $X$, let $\Phi\subseteq\Prop(\Lambda_{GML}(\Pow(X)))$ be
  maximally one-step consistent. We define $B\in\Baginfty(X)$ by
  $B(A)\ge n\iff\mge n.\,A\in\Phi$; it is easy to check that $B$ is
  well-defined and additive. To prove weak preservation of inverse
  limits, let $(X_n)$ be an $\omega$-cochain, let $X=\invlim X_n$, and
  let $(B_n\in\Baginfty(X_n))$ be coherent. Then define
  $B\in\Baginfty(X)$ pointwise by
\begin{equation*}
  B((x_n))=\min_{n\in\Nat}B_n(x_n),
\end{equation*}
noting that the sequence $(B_n(x_n))$ is decreasing by coherence. A
straightforward computation shows that $B$ induces $(B_n)$. Separation
is clear.

By the above and Thm.~\ref{thm:can-model}, all extensions of GML have
quasi-canonical multigraph models. While the technical core of the
construction is implicit in the work of Fine~\cite{Fine72} and
de~Caro~\cite{DeCaro88}, these authors were yet unaware of multigraph
semantics, and hence our result that \emph{GML is strongly complete
  over multigraphs} has not been obtained previously. 


The standard frame conditions for reflexivity, symmetry, and
transitivity (Expls.~\ref{expl:coalgml}.\ref{item:ax-gml}
and~\ref{expl:structure}.~\ref{item:gml-struct}) and arbitrary
combinations thereof are easily seen to be satisfied in the
quasi-canonical model constructed above. We point out that this
contrasts with Kripke semantics in the case of the graded version of
$S4$, i.e.\ GML extended with the reflexivity and transitivity axioms
of Expl.~\ref{expl:coalgml}.\ref{item:ax-gml}: as shown
in~\cite{FattorosiBarnabaCerrato88}, the complete axiomatisation of
graded modal logic over transitive reflexive Kripke frames includes
two rather strange combinatorial artefacts, which by the above
disappear in the multigraph semantics. The reason for the divergence
(which we regard as an argument in favour of multigraph semantics) is
that, while in many cases multigraph models are easily transformed
into equivalent Kripke models by just making copies of states, no such
translation exists in the transitive reflexive case (transitivity
alone is unproblematic).

Observe moreover that the above extends straightforwardly to
decription logics $\ALCQ(\RBox)$ with qualified number restrictions
and a role hierarchy $\RBox$ where roles may be distinguished as, in
any combination, transitive, reflexive, or symmetric. As shown
in~\cite{HorrocksEA99,KazakovEA07}, $\ALCQ(\RBox)$ is undecidable for
many $\RBox$, even when only transitive roles are considered. For
undecidable logics, completeness is in some sense the `next best
thing', as it guarantees if not recursiveness then at least recursive
enumerability of all valid formulas, and hence enables automatic
reasoning. Essentially, our results show that the natural
axiomatisation of \emph{$\ALCQ(\RBox)$ with transitive, symmetric and
  reflexive roles is strongly complete over multigraphs}, a result
which fails for the standard Kripke semantics.
\end{exa}


\begin{exa}[Failure of strong completeness of image-finite GML]
\label{expl:gml-finite}
Similarly to the case of image-finite Kripke frames, one can model an
image-finite version of graded modal logic coalgebraically by
exchanging the functor $\Baginfty$ for the \emph{finite multiset
  functor} $\Bag$, where $\Bag(X)$ consists of all maps $X\to\Nat$
with finite support. Of course, the resulting logic is non-compact and
hence fails to admit a canonical model. This is witnessed not only by
the same family of formulas as in the case of image-finite Kripke
semantics, which targets finiteness of the number of different
successors, but also by the set of formulas $\{\mge n.\,a\,\mid
n\in\Nat\}$, which targets finiteness of multiplicities. Analysing the
conditions of Thm.~\ref{thm:can-model}, we detect two violations: not
only does weak preservation of inverse limits fail, but there is also
no way to find an axiomatisation which is strongly one-step complete
over finite sets (again, consider sets $\{\mge n.\,\{x\}\mid
n\in\Nat\}$).
\end{exa}

\noindent Strong completeness of image-finite GML can be recovered by
slight adjustments to the syntax and semantics. We formulate a more
general approach, as follows.

\begin{exa}[Strong completeness of the logic of additive measures]
\label{expl:additive-measures}
We fix an at most countable commutative monoid~$M$ (e.g.\
$M=\Nat$). We think of the elements of $M$ as describing the measure
of a set of elements. To ensure compactness, we have to allow some
sets to have undefined measure. That is, we work with coalgebras for
the endofunctor $T_M$ defined by
\[ T_M(X) = \lbrace (\FA, \mu) \mid \FA \subseteq \Pow(X) \mbox{
  closed under disjoint unions}, \mu: \FA \to M \mbox{ additive}
\rbrace \] 
The modal logic of additive $M$-valued measures is given by the
similarity type $\Lambda_M = \lbrace E_m \mid m \in M \rbrace$ where
$E_m \phi$ expresses that $\phi$ has measure $m$, i.e.\
\begin{equation*}
  \Sem{E_m}_XB=\{(\FA,\mu)\in T_M(X)\mid B\in\FA,\mu(B)=m\}.
\end{equation*}
$\Lambda_M$ is clearly separating. The logic is axiomatised by the
following two axioms:
\begin{equation*}
 E_ma\modimpl\neg E_na\quad(n\neq m) 
\quad\textrm{and}\quad
 E_m(a\land b)\land E_n(a\land\neg b)\modimpl E_{m+n}a.
\end{equation*}
These axioms are strongly one-step complete over finite sets $X$: if
$\Phi\subseteq\Prop(\Lambda_M(\Pow(X)))$ is maximally one-step
consistent, then $(\FA,\mu)\modelsOS\Phi$ where $A\in\FA$ iff
$E_mA\in\Phi$ for some necessarily unique $m$, in which case
$\mu(A)=m$. Moreover, $T_M$ weakly preserves inverse limits
$X=\invlim X_n$, with finite $X_n$: a coherent family
$((\FA_n,\mu_n)\in T_M(X_n))$ is induced by
$(\FA,\mu)\in T_M(X)$, where $\FA=\{\pi_n^{-1}[B]\mid
n\in\Nat,B\in\FA_n\}$ and $\mu(\pi_n^{-1}[B])=\mu_n(B)$ is easily seen
to be well-defined and additive. Theorem~\ref{thm:can-model} now
guarantees existence of quasi-canonical models. A simple example is $M
= \Int / 2 \Int$, which induces a logic of even and odd.

For the case $M = \Nat$, we obtain a variant of graded modal logic
with finite multiplicities, where we code $\geq k. \phi$ as $\neg
\Lor_{0 \leq i < k} E_k \phi$. However, it may still be the case that
a state has a family of successor sets of unbounded measure, so that
undefinedness of the measure of the entire state space just hides an
occurrence of infinity. This defect is repaired by insisting that the
measure of the whole state space is finite at the expense of
disallowing the modal operator $E_0$ in the language, as follows.
\end{exa}



\newcommand{\GMLm}{\mathrm{GML}^-}
\begin{exa}[Strong completeness of finitely branching
  $\mathrm{GML}^-$]\label{expl:gmlm}

  To force the entire state space to have finite measure, we
  additionally introduce a \emph{measurability} operator~$E$,
  interpreted by $\Sem{E}B=\{(\FA,\mu)\mid B\in\FA\}$, and impose
  obvious axioms guaranteeing that measures on $X$ are defined on
  boolean subalgebras of $\Pow(X)$, in particular $E\top$ (i.e.\
  $\mu(X)$ is finite), and $E_na\modimpl Ea$. In order to achieve
  compactness, we now leave a bolt hole on the syntactical side and
  exclude the operator $E_0$. In other words, the syntax of $\GMLm$ is
  given by the similarity type
\( \Lambda_\GMLm = \lbrace E \rbrace \cup \lbrace E_n \mid n > 0
\rbrace \),
and we interpret $\GMLm$ over coalgebras for the functor 
$\Bag_M$ defined by
\[ \Bag_M(X) = \lbrace (\FA, \mu) \mid \FA \mbox{ boolean subalgebra
of $\Pow(X)$}, \mu: \FA \to \Nat \mbox{ additive} \rbrace.
\]
Separation is clear. 
%
%
The axiomatisation of $\GMLm$ is given by the axiomatisation of the
modal logic of additive measures, the above-mentioned axioms on $E$,
and the additional axiom 
\begin{gather*}
  E_n a\land E b \modimpl E_n (a\land b) \lor E_n(a\land\neg b) \lor
  \textstyle\Lor_{0<k<n} (E_k(a\land b)\land E_{n-k} (a\land\neg b))
\end{gather*}
which compensates for the absence of $E_0$. Strong one-step
completeness over finite sets and weak preservation of inverse limits
is shown analogously as in Expl.~\ref{expl:additive-measures}, so that
we obtain a \emph{strongly complete finitely branching graded modal
  logic $\GMLm$}. The tradeoff is that the operator $\geq k. \phi$
	is no longer expressible as $\neg \Lor_{0 \leq i < k} E_i
\phi$ in $\GMLm$ which only allows to formulate the implication
$\mge
1.\phi\modimpl\mge n.\,\phi$.
\end{exa}


\begin{exa}[Failure of strong completeness for PML over finitely
supported probability distributions]
Like image-finite graded modal logic, probabilistic modal logic as
introduced in Expl.~\ref{expl:coalgml}.\ref{item:pml} fails to be
compact, and violates the conditions of Thm.~\ref{thm:can-model} on
two counts, namely weak preservation of inverse limits and strong
one-step completeness over finite sets. The first issue is related to
image-finiteness, while the second is rooted in the structure of the
real numbers: e.g.\ the set $\{L_{1/2-1/n}a\mid n\in\Nat\}\cup\{\neg
L_{1/2}a\}$ is finitely satisfiable but not satisfiable. 
\end{exa}

\newcommand{\PMLe}{\mathrm{PML}_e}
\begin{exa}[Strong completeness of the logic of exact
probabilities]
In order to remove the above-mentioned failure of compactness, we
consider the fragment of probabilistic modal logic containing only
operators $E_p$ stating that a given event has probability exactly
$p$. (This is, of course, less expressive than the operators $L_p$ but
still allows reasonable statements such as that rolling a six on a die
happens with probability $1/6$.) Moreover, we require probabilities to
be rational and allow probabilities to be undefined, thus following
the additive measures approach as outlined above, where we consider a
subfunctor of $T_{\mathbb{Q}}$ defined by the requirement that the
whole set has measure~$1$. However, we are able to impose stronger
conditions on the domain $\FA\subseteq\Pow(X)$ of a probability
measure $P$ on $X$: we require that $X\in\FA$ and that $A,B\in\FA$,
$B\subseteq A$ imply $A-B\in FA$, which is reflected in the additional
axioms $E_1\top$ and $E_pa\land E_q(a\land b)\modimpl
E_{p-q}(a\land\neg b)$. It is natural that we cannot force closure
under intersection, as there is in general no way to infer the exact
probability of $A\cap B$ from the probabilities of $A$ and $B$.
Along the same lines as above, we now obtain quasi-canonical models,
and hence strong completeness and compactness, of the arising modal
logic of exact probabilities.

\end{exa}


\section{Conclusion}

\noindent We have laid out a systematic method of proving existence of
canonical models in a generic semantic framework encompassing a wide
range of structurally different modal logics. We have shown how this
method turns the construction of canonical models into an entirely
mechanical exercise where applicable, and points the way to obtaining
compact fragments of non-compact logics. As example applications, we
have reproved a number of known strong completeness result and
established several new results of this kind; specifically, the latter
includes strong completeness of the following logics.
\begin{sparitemize}
\item The modal logic of exact probabilities, with operators $E_p$
  `with probability exactly $p$'.
\item Graded modal logic over transitive reflexive multigraphs, i.e.\
  the natural graded version of $S4$, and more generally description
  logic with role hierarchies including transitive, reflexive, and
  symmetric roles and qualified number restrictions also on non-simple
  (e.g.\ transitive) roles.
\item The conditional logic $CK+\{\mi{ID},\mi{DIS}\}$, i.e.\ with the
  standard axioms of identity and disjunction, interpreted over
  conditional frames.
\end{sparitemize}
A number of interesting open problems remain, e.g.\ to find further
strongly complete variants of probabilistic modal logic or to
establish strong completeness of the full set of standard axioms of
default logic, Burgess' System C~\cite{Burgess81}, over the
corresponding class of conditional frames.

\bibliographystyle{myabbrv}\bibliography{coalgml}

\end{document}